\documentclass[aps,prl,notitlepage,amsmath,amssymb,amsfonts,superscriptaddress,twocolumn]{revtex4-2}
\makeatletter
\def\@hangfrom@section#1#2#3{\normalsize\@hangfrom{#1#2}#3}
\def\@hangfroms@section#1#2{\normalsize#1#2}
\makeatother

\usepackage{graphicx,soul}
\usepackage{dcolumn}
\usepackage{bm}
\usepackage{bbm} 

\usepackage{dsfont}
\usepackage{enumerate}

\usepackage[dvipsnames, x11names]{xcolor}
\usepackage[colorlinks=true]{hyperref}
\hypersetup{
    colorlinks=true,
    citecolor=blue,
    linkcolor=red,
    urlcolor=RoyalBlue,
    anchorcolor=Purple
}

\usepackage[percent]{overpic}
\usepackage{tabularx}
\usepackage{multirow}
\usepackage{CJK}
\usepackage{braket}

\newcommand{\bluefour}[1]{\textcolor{Blue4}{#1}}

\def\be{\begin{equation}}
\def\ee{\end{equation}}
\def\bea{\begin{eqnarray}}
\def\eea{\end{eqnarray}}

\begin{document}

\title{3D Ising criticality with Platonic lattice superconducting qubits}
\author{Liyang Sui}
\thanks{These authors contributed equally to this work.}
\affiliation{Zhejiang Province Key Laboratory of Quantum Technology and Device, School of Physics, Zhejiang University, Hangzhou 310027, China}
\author{Hong-Hao Song}
\thanks{These authors contributed equally to this work.}
\affiliation{International Center for Quantum Materials, School of Physics, Peking University, Beijing 100871, China}
\author{Sainan Huai}
\thanks{These authors contributed equally to this work.}
\affiliation{Tencent Quantum Laboratory, Tencent, Shenzhen 518057, China}
\author{Yufan Li}
\affiliation{Zhejiang Province Key Laboratory of Quantum Technology and Device, School of Physics, Zhejiang University, Hangzhou 310027, China}
\author{Zhiwen Zong}
\affiliation{Tencent Quantum Laboratory, Tencent, Shenzhen 518057, China}
\author{Kunliang Bu}
\affiliation{Tencent Quantum Laboratory, Tencent, Shenzhen 518057, China}
\author{Xiaopei Yang}
\affiliation{Tencent Quantum Laboratory, Tencent, Shenzhen 518057, China}
\author{Xingrui Liu}
\affiliation{Zhejiang Province Key Laboratory of Quantum Technology and Device, School of Physics, Zhejiang University, Hangzhou 310027, China}
\author{Wenyan Jin}
\affiliation{Zhejiang Province Key Laboratory of Quantum Technology and Device, School of Physics, Zhejiang University, Hangzhou 310027, China}
\author{Bowen Chen}
\affiliation{Zhejiang Province Key Laboratory of Quantum Technology and Device, School of Physics, Zhejiang University, Hangzhou 310027, China}
\author{Xutao Zhang}
\affiliation{Zhejiang Province Key Laboratory of Quantum Technology and Device, School of Physics, Zhejiang University, Hangzhou 310027, China}
\author{Jianlan Wu}
\affiliation{Zhejiang Province Key Laboratory of Quantum Technology and Device, School of Physics, Zhejiang University, Hangzhou 310027, China}

\author{Yicong Zheng}
\affiliation{Tencent Quantum Laboratory, Tencent, Shenzhen 518057, China}

\author{Shengyu Zhang}\thanks{shengyzhang@tencent.com}
\affiliation{Tencent Quantum Laboratory, Tencent, Shenzhen 518057, China}

\author{Gang v.~Chen}
\thanks{chenxray@pku.edu.cn}
\affiliation{International Center for Quantum Materials, School of Physics, Peking University, Beijing 100871, China}
\affiliation{Collaborative Innovation Center of Quantum Matter, 100871, Beijing, China}
\affiliation{Beijing Key Laboratory of Quantum Devices, Peking University, Beijing 100871, China}
\author{Yi Yin} \thanks{yiyin@zju.edu.cn}
\affiliation{Zhejiang Province Key Laboratory of Quantum Technology and Device, School of Physics, Zhejiang University, Hangzhou 310027, China}
\date{\today}

\begin{abstract}
The three-dimensional(3D) Ising model is a foundational model in statistical physics
and critical phenomena, yet its analytical intractability has long impeded the
precise determination of universal critical exponents. While the high-precision
estimates have been obtained through classical numerical methods and conformal
bootstrap techniques, a direct quantum simulation of the 3D Ising criticality
remains challenging, requiring the nontrivial connectivity, sufficient system
size, and high spectral resolution. In this work,
assisted by the state-operator correspondence of the conformal field theory,
we perform a digital quantum simulation of the 3D Ising critical exponents using a
multiply-connected 9-qubit superconducting quantum processor with a Platonic lattice geometry.
Employing an extended variational quantum eigensolver equipped with a
phase-based loss function, we variationally prepare the low-energy
eigenstates of the transverse-field Ising model on a cubic Platonic lattice
encoded in an 8-qubit register. The four lowest eigenenergies are extracted
via Fourier-transform analysis and high-precision numerical fitting,
agreeing with the exact diagonalization values up to ${\pm 0.001}$.
The resulting scaling dimension ${\Delta_\epsilon = 1.5850}$ and
critical exponent ${\nu = 0.7067}$ match well with the theory.
Our results establish a scalable pathway for
studying the critical properties of quantum
many-body physics using near-term quantum devices.
\end{abstract}

\maketitle

\noindent\bluefour{\it Introduction.}---Phase transitions and critical
phenomena are among the most important and challenging subjects in modern physics.
Among the numerous models in this field, the three-dimensional (3D)
Ising model plays a particularly important role. Despite its
simple formulation, it has denied an exact analytical solution
for decades, in sharp contrast to the two-dimensional case solved
by Onsager~\cite{PhysRev.65.117}. This long-standing difficulty
has motivated extensive efforts from multiple directions~\cite{PELISSETTO2002549},
including searches for field theory~\cite{PhysRevD.96.036016},
large scale numerical simulations~\cite{PhysRevB.82.174433,PhysRevE.97.043301},
conformal bootstrap~\cite{RevModPhys.91.015002,Simmons-Duffin2017,PhysRevD.86.025022,PhysRevD.80.045006,Kos2016},
and fuzzy sphere regularization~\cite{PhysRevX.13.021009}.
These studies have greatly deepened our understanding of the
3D Ising universality class and, more broadly, advanced the
modern understanding of critical phenomena. Quantum
simulations access to 3D Ising criticality, however,
remains highly challenging,
due to the combined demands of three-dimensionality, sufficiently
large system sizes, and precise probing of the critical
regime~\cite{sun2026experimentalobservationconformalfield}.

After decades of theoretical exploration, it is widely believed
that the infrared critical point of the 3D Ising model is
described by a conformal field theory (CFT)~\cite{Polyakov:1970xd,PhysRevX.13.021009}.
Within the CFT framework, the state-operator correspondence
relates the scaling operators in the flat space to the energy eigenstates
of the theory quantized on $S^2 \times \mathbb{R}$, thereby
providing a direct route to extracting the CFT data from the
finite-size simulations~\cite{Cardy_1984,Cardy_1985}.
Capturing the underlying CFT requires that the lattices maximally preserve
the $O(3)$ symmetry of the sphere, making the
Platonic lattices a natural choice.
Exact diagonalization (ED) studies indicate that the low-energy
spectrum of the transverse field Ising model (TFIM) on Platonic
lattices agrees well with the low-lying spectrum of the 3D Ising CFT,
making the Platonic lattices a promising platform for the
quantum simulation of 3D Ising
criticality~\cite{10.21468/SciPostPhys.15.6.243,wu2026qubitdiscretizationsd3conformal}.

Critical exponents are universal quantities characterizing
continuous phase transitions, governing the scaling behavior
of observables near criticality. Within the CFT framework,
these exponents are directly related to the scaling dimensions
of a few primary field operators~\cite{Cardy_1996,Henkel1999qx},
which are in turn encoded in the few lowest energy levels of the
TFIM on Platonic lattices. This feature makes the simulation of
critical exponents particularly efficient, since an acceptable
accurate estimate may be obtained without requiring large
system sizes (see Supplementary Information Section I).
Together with the high controllability
and precision of modern quantum simulators, this opens a
realistic route to probing the 3D Ising critical exponents.

\begin{figure}[tp]
    \centering
    \includegraphics[width=1.0\linewidth]{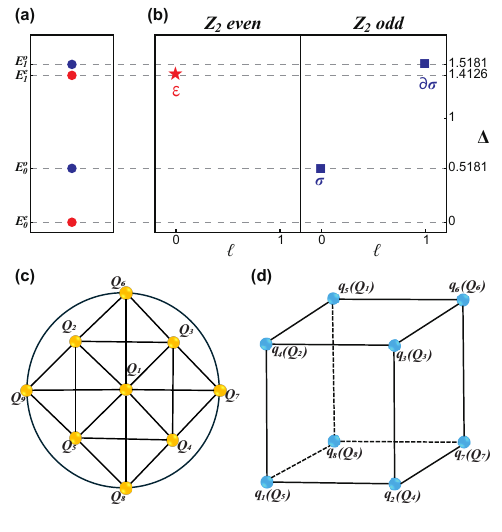}
    \caption{State-operator correspondence and device schematic.
    (a) The four lowest renormalized energy levels of the
    TFIM on sphere. Red circles denote the levels in the $\mathbb{Z}_2$ even
    sector, while blue circles denote the levels in the $\mathbb{Z}_2$ odd sector.
    (b) Low-lying spectrum of the 3D Ising CFT. The red star marks
    the scaling dimension of the primary field $\epsilon$, while
    the blue squares mark the scaling dimensions of the primary
    field $\sigma$ and its descendant $\partial \sigma$.
    (c) Schematic of the L9 processor, showing nine qubits with
    a tunable coupler integrated into every inter-qubit link (solid black line).
    (d) Cubic lattice mapped onto the L9 chip: each vertex
    corresponds to a physical qubit, as indicated by the bracketed labels.
    }
    \label{fig1}
\end{figure}

In this work, we report the digital quantum simulation of the
critical exponents of the 3D Ising model using
a superconducting multi-qubit processor~\cite{BarendsPRL13,Krantz19}.
By implementing an
extended variational quantum eigensolver (VQE) protocol equipped
with a phase-based loss function, we variationally prepare
low-energy eigenstates of the TFIM
on a cubic Platonic lattice encoded in an 8-qubit register.
The corresponding eigenenergies are extracted with high spectral
resolution via Fourier-transform analysis followed by high-precision
numerical fitting. The phase-based loss function naturally
treats ground and excited states on equal
footing, enabling the simultaneous determination of four
low-lying energy levels that are required for obtaining the scaling dimensions
$\Delta_\epsilon$ of the 3D Ising CFT.
We experimentally demonstrate a complete variational workflow
on L9, a 9-qubit superconducting processor whose qubit
connectivity matches that shown in Fig.~\ref{fig1}c.
This includes the on-device state preparation and circuit
optimization, followed by the spectral analysis of the
optimized output states. The L9 architecture
features native
 long-range couplings within its multi-connected
qubit lattice.
Specifically, a subset of 8 qubits in L9 realizes couplings
that naturally match the cubic interaction graph of the Platonic lattice.
We further validate the generality of the method by
implementing the simulation on a linear chain of 8 qubits,
confirming that the results are robust against the underlying
hardware connectivity. This work establishes a viable pathway
for probing 3D critical phenomena using current noisy
intermediate-scale quantum (NISQ) devices~\cite{Preskill18}.

\medskip

\noindent\bluefour{\it Theoretical Framework.}---
The key theoretical foundation of our approach is the state-operator
correspondence.
For a 3D CFT, radial quantization maps local scaling operators
in flat space $\mathbb{R}^3$ to the energy eigenstates of the theory
quantized on $S^2 \times \mathbb{R}$~\cite{Cardy_1985}.
As a result, the scaling dimensions $\Delta_i$ of the local operators
are directly related to the energy spectrum on the sphere,
so the latter provides direct access to the CFT data.
Figure~\ref{fig1}b shows the
low-lying spectrum of the 3D Ising CFT, including the primary
fields $\epsilon$ and $\sigma$ as well as the descendant $\partial\sigma$.
Figure~\ref{fig1}a shows the
corresponding renormalized energy levels of the TFIM on the sphere,
defined as
$E_i^{e,o} = (\tilde{E}_i^{e,o}-\tilde{E}_0^e)R$,
where $\tilde{E}_i^{e,o}$ denotes the $\mathbb{Z}_2$ even and
$\mathbb{Z}_2$ odd energy eigenvalues of the lattice Hamiltonian
and $R$ is the radius of the sphere.
In the continuum limit, these renormalized energy levels equal
to the scaling dimensions of the corresponding CFT operators exactly.

In particular, 3D Ising critical exponents are encoded in the
primary field $\sigma$ and the primary field $\epsilon$.
Once $\Delta_\sigma$ and $\Delta_\epsilon$ are extracted, the
critical exponents follow from~\cite{Cardy_1996,Henkel1999qx}
\begin{equation}
\eta = 2\Delta_\sigma - 1,\qquad
\nu = \frac{1}{3-\Delta_\epsilon}.
\end{equation}
Therefore, determining only a few low-lying energy levels is
sufficient to infer the critical exponents of the 3D Ising
universality class.

The lattice implementation of this idea requires a discretization
that maximally preserves the O(3) symmetry of the
sphere.
Natural candidates are Platonic lattices, including the octahedron
and the cube, which form a dual pair and possess the full
octahedral symmetry $O_h$, as well as the icosahedron and the
dodecahedron, which form another dual pair with the full icosahedral
symmetry $I_h$.
These symmetry groups are discrete subgroups of $O(3)$, therefore
provide a natural setting for approximating the low-energy
spectrum of the continuum theory.
Another important ingredient is the special structure of the
3D Ising CFT itself.
Its low-lying spectrum is sparse, and the critical exponents are
determined by only the two lowest scaling dimensions.
These make it possible to obtain acceptable results even on
lattices with a modest number of sites.

We are now led to consider the TFIM on the Platonic lattices,
\begin{equation}
H_\mathrm{p} = - J \sum_{\langle ij\rangle} \sigma_i^z \sigma_j^z - h \sum_i \sigma_i^x .
\label{eq:Hamiltonian}
\end{equation}
Here $\sigma_i^{x,y,z}$ are Pauli matrices, and $\langle ij\rangle$
denotes nearest-neighbor pairs on the Platonic lattice. The parameter $J$
denotes the Ising coupling, and $h$ denotes the transverse-field strength.
The Hamiltonian
is invariant under the global $\mathbb{Z}_2$ symmetry generated
by $U=\prod_i \sigma_i^x$, which classifies the eigenstates
into $\mathbb{Z}_2$ even and $\mathbb{Z}_2$ odd sectors.

Although the relation between scaling dimensions and energies,
$\Delta_i^{e,o} = R(\tilde{E}_i^{e,o}-\tilde{E}_0^e)$, is exact
only in the continuum limit, the ratios are still universal even
for finite system sizes fortunately. We therefore need an
scaling factor to match the lattice energy spectrum
to the CFT data. In addition, a criterion is required to
locate the critical transverse field $h_c$. In our approach,
both are fixed from the $\mathbb{Z}_2$ odd sector. The factor
is set by $R=\Delta_{\sigma}/(\tilde{E}_0^o-\tilde{E}_0^e)$,
and the critical field is determined from the condition ${E_1^o-E_0^o=1}$.
Once these are done, the first nontrivial level in the $\mathbb{Z}_2$
even sector yields the scaling dimension $\Delta_\epsilon$.
More details are provided in the Supplementary Information Section I.

Our proposal only requires the four lowest energy levels and
a modest number of qubits, which makes it particularly well
suited to quantum simulation. In finite size simulations,
only the lowest few energy levels can be expected to reliably
capture the CFT data, while higher levels are generally more
strongly affected by finite-size effects. From the experimental
perspective, there are two additional challenges. First,
resolving a larger number of excited states is increasingly
difficult. Second, as the number of sites increases, resolving
the energy spectrum becomes more demanding even for the
low-lying levels. The present scheme strikes a favorable
balance between these constraints and can still provide
estimates with acceptable accuracy.

\medskip

\noindent\bluefour{\it Experimental Protocol and Results.}---
We simulate the cubic Platonic lattice using an 8-qubit register
matching its 8 vertices, governed by a TFIM Hamiltonian $H_\mathrm{p}$ with Ising coupling $J$.
The critical transverse field is $h_c/J = 2.46$, determined by
$E_1^o - E_0^o = 1$ (see Supplementary Information Section~I).
We simulate directly at this numerically determined critical point
to extract the four lowest eigenenergies required for critical exponent
calculations. To this end, we employ the multi-connected superconducting
processor L9, wherein a selected 8-qubit subset natively realizes the
12-edge interaction graph of the cubic lattice.

Figure~\ref{fig1}c presents a schematic of the L9 processor, which
consists of nine frequency-tunable transmon
qubits ($Q_1$--$Q_9$)~\cite{BarendsPRL13,BunpjQuantInf25}.
Based on flip-chip packaging, the device integrates the qubits
on the top wafer with the control lines and resonator-based
readout structures on the bottom wafer, yielding a compact
two-dimensional layout. Here, $Q_1$ acts as a central hub coupled
to all eight peripheral qubits ($Q_2$--$Q_9$) through individual
Josephson-junction-based tunable couplers, while the peripheral
qubits are further interconnected via an additional sixteen
tunable couplers, resulting in a total of 24 independently
tunable couplers in the full 9-qubit architecture. The qubits
are coupled via floating tunable couplers that suppress
residual $ZZ$ crosstalk~\cite{SetePRApplied2021}. The cross
structure of the coupler in the multi-connected system shown in
Fig.~\ref{fig1}c is implemented by a high-performance
tantalum air bridge~\cite{BunpjQuantInf25}. More details
about the chip design and fabrication will be reported in
another companion article~\cite{L9}.
Each transmon exhibits a $4$-$5$~GHz idle frequency,
$\approx -200$~MHz anharmonicity, average $T_1 = 57.3~\mu\text{s}$
($37.9$--$85.6~\mu\text{s}$), and $T_2^\text{echo} = 26.8~\mu\text{s}$
($10.3$--$47.3~\mu\text{s}$).
Full parameters are in Supplementary Table~IV.

The cube topology---8 vertices and 12 edges (Fig.~\ref{fig1}d)---can
therefore be encoded directly on the L9 chip.
In our simulation we restrict the active register to
8 selected qubits $Q_1$--$Q_8$ (excluding $Q_9$).
The one-to-one mapping between circuit qubits $\{q_i\}$
and physical qubits $\{Q_j\}$ is illustrated in Fig.~\ref{fig1}d.
Under this mapping, the native couplings among $Q_1$--$Q_8$
realize the 12-edge connectivity graph of the cube.
The floating tunable couplers support high-fidelity, fast controlled-$Z$ (CZ)
gates across this multi-connected architecture.
For the 8-qubit register, the average single-qubit gate
fidelity, determined via Clifford randomized benchmarking (RB), is 99.93\% (ranging
from 99.91\% to 99.95\%); the two-qubit CZ-gate fidelities
range from 98.36\% to 99.94\%, with a mean of 99.46\%.
As shown in the Supplementary Information II, all gate-control
and measurement operations are generated by a multi-channel
arbitrary waveform generator (AWG) \cite{ZhangIEEE2021}.
Full gate performance metrics are reported in Supplementary Table IV and V.

To compute the low-lying eigenenergies of the 8-qubit
system in Eq.~(\ref{eq:Hamiltonian}), we implement an extended VQE protocol
with an 8-qubit parameterized circuit. A phase-based
loss function is employed to steer the circuit toward
quantum states that approximate a targeted subset of
low-energy eigenstates. The corresponding eigenenergies
are subsequently extracted with great accuracy through
Fourier-transform analysis followed by high-precision
numerical fitting. The working
principle of the protocol and the obtained results
are presented in the following paragraphs.

\begin{figure*}[tp]
    \centering
    \includegraphics[width=1.0\linewidth]{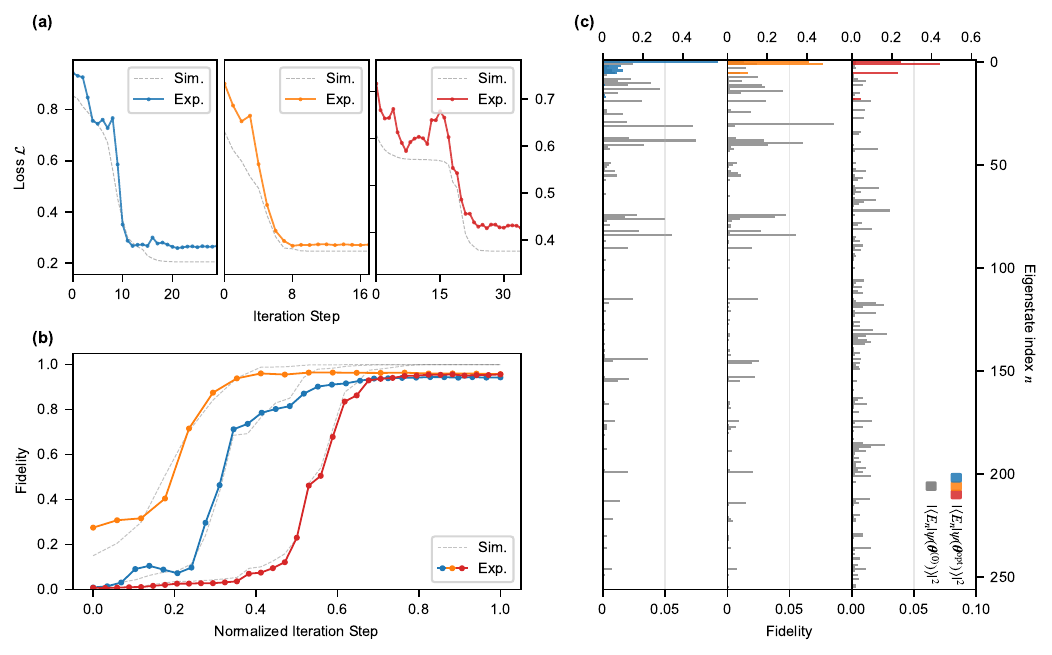}
    \caption{Variational optimization of quantum states
    for three different initial configurations.
    (a) Loss function $\mathcal{L}$ versus optimization iteration step for
    initial states $\ket{00000001}$, $\ket{00000010}$,
    and $\ket{00000101}$, with corresponding parameter of evolution times
    $\tau = 13.2$, $12.7$, $16.7$, respectively. The dashed grey curves denote numerical
    simulation; solid circles connected by lines are experimental data, shown in blue,
    orange, and red. The loss decays and converges approximately
    at steps 10, 8, and 24, respectively.
    (b) State fidelity with respect to the final converged state as a
    function of normalized iteration step (mapped to $[0,1]$ for
    comparison). The same color scheme and line styles as in (a) are used.
    (c) Spectral weight distribution $|c_n|^2$ of the simulated initial
    $\ket{\psi(\bm{\theta}^{(0)})}$ (grey) and final
    converged states $\ket{\psi(\bm{\theta}^\mathrm{opt})}$,
    projected onto the exact eigenbasis obtained from full diagonalization of the Hamiltonian,
    shown for the same three configurations (blue, orange, red).
    }
    \label{fig2}
\end{figure*}

For the 8 qubits ($Q_1$--$Q_8$) of the L9 chip---configured to match the vertex connectivity of the cube---we employ a
hardware-efficient~\cite{KandalaNature17}, two-layer parameterized circuit
as the variational ansatz (see Methods and Supplementary
Fig.~S3(a)). Each layer comprises parallel single-qubit $R_Y$ rotations
and a fixed pattern of CNOT gates generating entanglement along the twelve
cube edges. This ansatz is applied to computational basis states
($\ket{S}={\otimes _{i=1}^8}\ket{s_i}$ with $s_i \in \{0,1\}$),
rapidly prepared via single-qubit $X$ gates to establish different experimental
initial conditions. Taking $\ket{00000001}$, $\ket{00000010}$,
and $\ket{00000101}$ as three exemplary initial states $\ket{\psi_0}$,
the initial parameters $\bm{\theta}^{(0)}=2\pi\mathbf{1}_{16}$
yield the reference state $\ket{\psi({\bm{\theta}^{(0)}})}=\ket{\psi_0}$.

To optimize $\bm{\theta}$ via classical-quantum hybrid iteration,
we aim to evolve these initial states toward the target states.
Setting $\hbar=1$, we define a phase-based loss function for a chosen
evolution time $\tau$ (discussed later):
\begin{equation}
\mathcal{L}(\bm{\theta};\tau) = 1 - \bigl|\langle\psi(\bm{\theta})|e^{-iH_{\mathrm{p}}\tau}|\psi(\bm{\theta})\rangle\bigr|.
\label{eq:loss}
\end{equation}
If $\ket{\psi(\bm{\theta})}$ is an eigenstate of $H_{\mathrm{p}}$,
the expectation value retains a strict unit modulus,
identically vanishing the loss ($\mathcal{L}=0$);
conversely, state superpositions degrade this modulus, rendering the loss
strictly positive ($\mathcal{L}>0$).
Unlike standard VQE~\cite{PeruzzoNatCommun14,TillyPhysRep22},
which minimizes the energy $\langle\psi(\bm{\theta})|H_{\mathrm{p}}|\psi(\bm{\theta})\rangle$,
Eq.~(\ref{eq:loss}) treats all eigenstates equally.
This naturally enables targeting excited states without
explicit orthogonalization or subspace penalties~\cite{HiggottQuantum19,
Gocho2023,Nakanishi2019}.

For three distinct $(\ket{\psi_0}, \tau)$ configurations,
the numerical loss $\mathcal{L}(\bm{\theta}^{(i)};\tau)$
monotonically decays and plateaus at small nonzero values
(dashed curves, Fig.~\ref{fig2}a). The optimization terminates
when either the relative change in the loss function or the
gradient norm drops below the tolerance threshold $10^{-8}$,
stably converging within 35 iterations.
Relative to the final converged state $\ket{\psi_\mathrm{con}}$,
the simulated state fidelity
$f_\mathrm{sim}^{(i)}= \bigl|\langle\psi_\mathrm{con}|\psi_\mathrm{sim}^{(i)}\rangle\bigr|^2$
monotonically grows and saturates at convergence (dashed curves,
Fig.~\ref{fig2}b). For direct comparison across configurations,
the iteration axis in Fig.~\ref{fig2}b is normalized to the unit interval $[0, 1]$.

Through ED, we calculate the spectral weights
$|c_n|^2$ ($\ket{\psi}= \sum_{n}c_n\ket{\psi_n}$) of the initial and
converged states in the 256-dimensional eigenbasis (Fig.~\ref{fig2}c).
While initial states are broadly delocalized, converged states become
sharply localized, predominantly comprising a few low-energy eigenstates.
This residual distribution explains the finite, nonzero converged loss.
Although the loss function treats all eigenstates equally, optimization
from generic initial states inherently favors low-lying eigenstates
(see Supplementary Information Section IV).

As the variational states are real superpositions,
we characterize them via single-shot measurements, avoiding full
quantum state tomography (QST) (Methods). The experimental loss
$\mathcal{L}_\mathrm{exp}(\bm{\theta}^{(i)};\tau)$, calculated
from corrected states $\ket{\psi_\mathrm{exp}^{(i)}}$, decreases
similarly to simulations (solid circles, Fig.~\ref{fig2}a).
The experimental fidelity
$f_\mathrm{exp}^{(i)}= \bigl|\langle\psi_\mathrm{con}|\psi_\mathrm{exp}^{(i)}\rangle\bigr|^2$
stably converges to 94\%--96\% (solid circles, Fig.~\ref{fig2}b),
limited primarily by inherent experimental errors. To save quantum
resources, we use classically computed gradients, though experimental
gradients are also viable (see Supplementary Information Section III
and Fig.~S5).

\begin{figure*}[tp]
    \centering
    \includegraphics[width=1.0\linewidth]{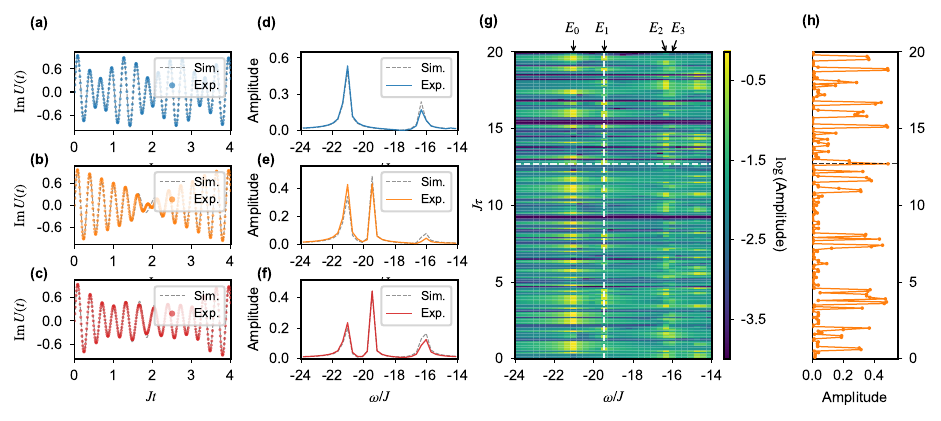}
    \caption{Time evolution and spectral analysis for the three optimized states.
    (a-c) Imaginary part of the expectation value
    $\text{Im}\langle \psi_{\text{exp}}^{\text{opt}} | e^{-iH_{\text{p}}t} | \psi_{\text{exp}}^{\text{opt}} \rangle$
    versus time $t$ for the final states prepared
    from the initial configurations $|\psi_0\rangle = |00000001\rangle$ (a),
    $|00000010\rangle$ (b), and $|00000101\rangle$ (c), each with its
    corresponding evolution time $\tau$ used in the optimization. The time series
    is computed over a window $Jt \in [0, 20]$; panels (a-c) show an
    enlarged view of the short-time region $Jt \in [0, 4]$. Dashed grey curves
    are numerical simulations; solid circles (blue, orange, red)
    are calculated from the experimentally measured states.
    (d-f) Discrete Fourier transform (DFT) amplitude of the full time-domain signals
    (up to $Jt=20$) shown in (a-c), plotted against angular frequency $\omega/J$.
    Grey dashed lines are the simulated DFT spectra;
    colored solid lines correspond to the experimental data.
    Arranged in order of energy from high to low, (d-f) respectively
    contain peaks of the $\{0,2,3\}$, $\{0,1,3\}$ and $\{0,1,3\}$ energy levels.
    (g) Fourier-transform spectrum for the initial state $|00000010\rangle$.
    The map shows the amplitude (on a log scale) versus normalized angular
    frequency $\omega/J$ (horizontal axis) and normalized evolution time
    $J\tau$ (vertical axis), obtained from the time-domain signal of
    the optimized state at each $\tau$. Arrows indicate the vertical bright
    strips corresponding to the four lowest eigenenergies $E_0$--$E_3$.
    The horizontal dashed line corresponds to the $\tau$ used in (e),
    and the vertical dashed line corresponds to the amplitude trace at $E_1$ shown in (h).
    The black dashed line in (h) marks the selected $\tau$ that yields
    a large amplitude for the $E_1$ peak.
    }
    \label{fig3}
\end{figure*}

Using these converged experimental states
$\ket{\psi_\mathrm{exp}^{\mathrm{opt}}}=\ket{\psi_\mathrm{exp}(\bm{\theta}^\mathrm{opt})}$,
we compute the time-dependent expectation
$U(t)=\langle\psi_\mathrm{exp}^{\mathrm{opt}}|e^{-iH_{\mathrm{p}}t}|\psi_\mathrm{exp}^{\mathrm{opt}}\rangle$,
whose imaginary part is
$\mathrm{Im}\,U(t)= -\!\sum_{n}\!|c_n|^2\sin(E_n t)$.
Plotted in dimensionless units ($Jt$, $\omega/J$),
Figs.~\ref{fig3}a--c show the short-time dynamics of
$\mathrm{Im}\,U(t)$ for $Jt\in[0,4]$, where the experimental data
agree well with the simulated dynamics (dashed lines).

Because the optimized states primarily consist of a few
low-lying eigenstates, we can extract the eigenenergies $E_n$
via spectral analysis of $\operatorname{Im} U(t)$. Applying a
discrete Fourier transform (DFT) to the signal over $Jt \in [0, 20]$
yields the experimental spectra (Figs.~\ref{fig3}d--f).
Consistent with simulations, these spectra exhibit peaks corresponding
to energy-level sets such as $\{0,2,3\}$, $\{0,1,3\}$, and $\{0,1,3\}$
for the three initial states. For $|\psi_0\rangle = |00000010\rangle$,
a simulated 2D map of the Fourier amplitude (log scale) versus
$\omega/J$ and $J\tau$ (Fig.~\ref{fig3}g) reveals bright vertical stripes
at the eigenenergies (e.g., $E_1$), enabling the estimation of $E_0$
through $E_3$ without prior knowledge. The frequency resolution
$\Delta(\omega/J) \approx 0.31$ introduces an energy uncertainty of
$\mathcal{O}(10^{-1})$, which is sufficient for preliminary spectral
identification. Furthermore, this map guides the selection of $\tau$
to maximize the signal amplitude for target energies; for instance,
to target $E_1$, we choose $J\tau = 12.7$ at a local maximum
(Fig.~\ref{fig3}h). Conversely, dark horizontal lines indicate $\tau$
values where the optimization converges to higher-energy eigenstates.
Similar $\tau$-dependent analyses for other initial states are detailed
in Supplementary Information Section~IV.

Prior to extracting eigenenergies from the Fourier spectra,
we examine the symmetry properties of the Hamiltonian\textquoteright s eigenstates.
For an optimized variational state $\ket{\psi^{\mathrm{opt}}}$,
we introduce its dual state
$\ket{\psi^\mathrm{h}}={\otimes _i}\hat{X}_i\ket{\psi^{\mathrm{opt}}}$,
where ${\otimes _i}\hat{X}_i$ denotes the global bit-flip operator
acting on all 8 qubits (equivalently $\prod_i \sigma_i^x$).
Since $[H_\mathrm{p},{\otimes _i}\hat{X}_i]=0$, the eigenstates can be
labeled by the eigenvalue $\pm1$ of ${\otimes _i}\hat{X}_i$, corresponding
to $\mathbb{Z}_2$ even and $\mathbb{Z}_2$ odd parity sectors.
Experimentally, $\ket{\psi^\mathrm{h}}$ is prepared by appending
single-qubit $X$ gates to the circuit that generate
$\ket{\psi^{\mathrm{opt}}}$. For the three configurations considered,
both the original state $\ket{\psi_\mathrm{exp}^{\mathrm{opt}}}$ and its dual
$\ket{\psi_\mathrm{exp}^{\mathrm{h}}}$
have been prepared and measured.
Using the two experimentally measured states, we construct the time-dependent signals
$U(t)=\bra{\psi_\mathrm{exp}^{\mathrm{opt}}}e^{-iH_{\mathrm{p}}t}\ket{\psi_\mathrm{exp}^{\mathrm{opt}}}$
and
$U^h(t)=\bra{\psi_\mathrm{exp}^{\mathrm{opt}}}e^{-iH_{\mathrm{p}}t}\ket{\psi_\mathrm{exp}^{\mathrm{h}}}$.
Half the sum and difference of their imaginary parts yield,
respectively, the $\mathbb{Z}_2$ even and $\mathbb{Z}_2$ odd parity channel signals:
\begin{align}
S_{+}(t) &\equiv \tfrac{1}{2}\bigl[\mathrm{Im}\,U(t)+\mathrm{Im}\,U^{\mathrm{h}}(t)\bigr]
    = -\!\sum_{n\in\mathrm{even}}\!|c_n|^2\sin(E_n t), \label{eq:Splus}\\
S_{-}(t) &\equiv \tfrac{1}{2}\bigl[\mathrm{Im}\,U(t)-\mathrm{Im}\,U^{\mathrm{h}}(t)\bigr]
    = -\!\sum_{n\in\mathrm{odd}}\!|c_n|^2\sin(E_n t). \label{eq:Sminus}
\end{align}

\begin{figure}[tp]
    \centering
    \includegraphics[width=1.0\linewidth]{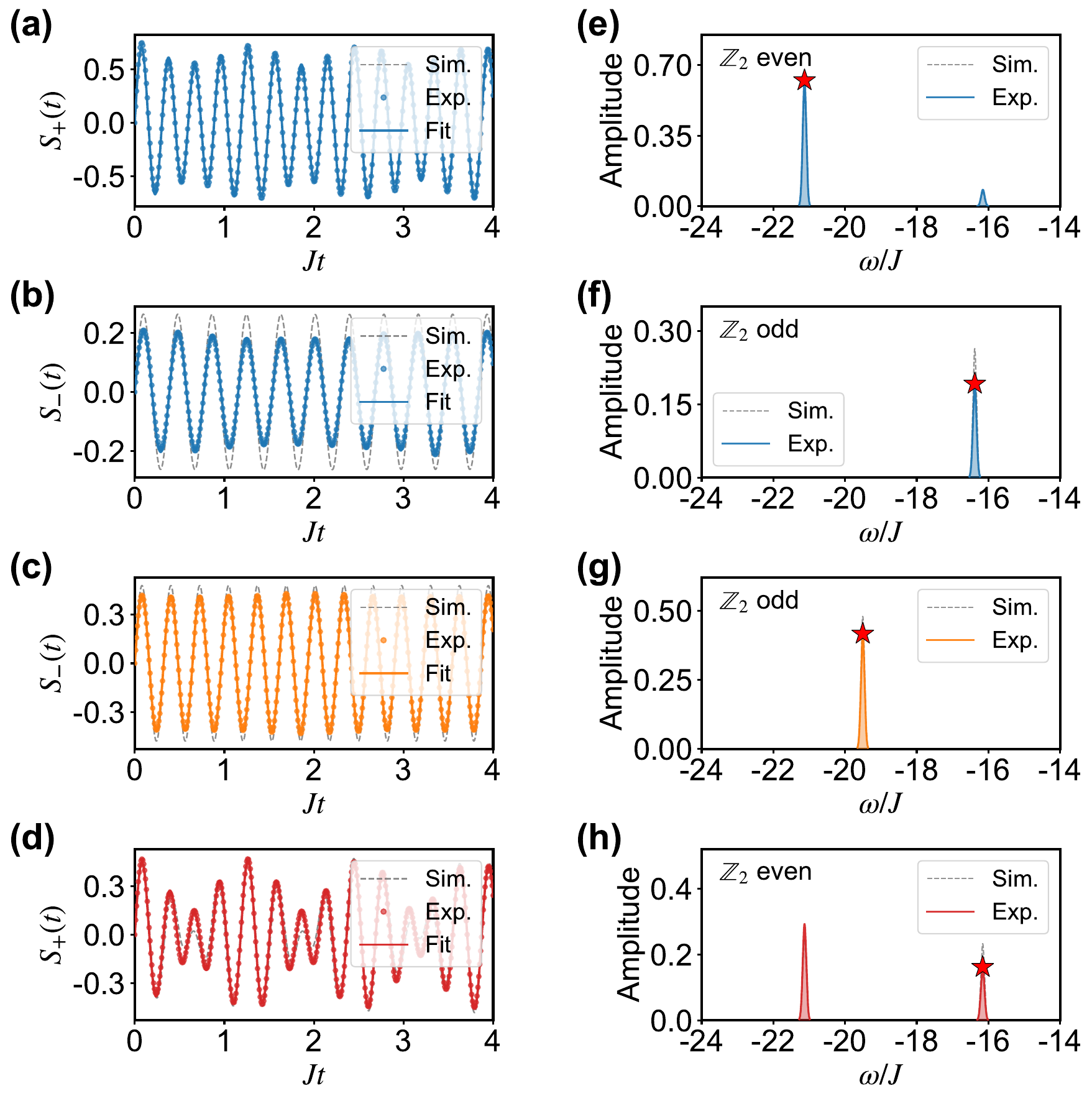}
    \caption{Parity-resolved time-domain signals and spectral identification of eigenenergies.
    (a-d) Imaginary part of the parity-resolved time-domain signals for the three initial configurations.
    (a, b) Even- (a) and odd-channel (b) signals $S_+(t)$ and $S_-(t)$ for the state prepared from $\ket{00000001}$ ($J\tau = 13.2$).
    (c) Odd-channel signal $S_+(t)$ for the state from $\ket{00000010}$ ($J\tau = 12.7$).
    (d) Even-channel signal $S_-(t)$ for the state from $\ket{00000101}$ ($J\tau = 16.7$).
    In all panels, grey dashed curves show numerical simulations;
    solid colored circles represent experimental data; solid colored lines are fitting curves to the experimental points.
    (e-h) Amplitude spectra obtained from multi-component sinusoidal fitting of the experimental signals in (a-d),
    plotted against angular frequency $\omega/J$.
    (e) Spectrum of the even-channel signal (a), showing peaks at eigenvalues $\{\tilde{E}_0^{e},\tilde{E}_1^{e}\}$.
    (f) Spectrum of the odd-channel signal (b), showing a peak at eigenvalue $\{\tilde{E}_1^{o}\}$.
    (g) Spectrum of the odd-channel signal (c), retaining eigenvalue $\{\tilde{E}_0^{o}\}$.
    (h) Spectrum of the even-channel signal (d), retaining eigenvalues $\{\tilde{E}_0^{e},\tilde{E}_1^{e}\}$.
    All fitted peaks have an artificially set Gaussian broadening of 0.05, exceeding $100\times$ the fitting error.
    Red stars mark the extracted eigenenergies and the corresponding state fidelities (amplitudes) for the four lowest levels.
    }
    \label{fig4}
\end{figure}

\medskip
\noindent\textit{Remark.}---Once the $\mathbb{Z}_2$ even and $\mathbb{Z}_2$ odd components of the time-domain signal are separated,
the energy levels $\{0,1,2,3\}$ can be clearly assigned to the extracted energy eigenvalues $\{\tilde{E}_i^{e,o}\}$ within each symmetry sector.
All subsequent expressions for scaling dimensions and critical exponent calculations are based on this symmetry-resolved identification.
\medskip

Figure~\ref{fig4}a--b display, for the first configuration, the $\mathbb{Z}_2$ even
and $\mathbb{Z}_2$ odd parity time-domain signals $S_{+}(t)$ and $S_{-}(t)$.
Then starting from their DFT spectra, we perform a global multi-frequency sinusoidal fit on the $S_{\pm}(t)$ signals using the form
\begin{equation}
f(t) = -\sum_{i}A_i \sin(\omega_i t + \varphi_i) + C,
\label{eq:multifit}
\end{equation}
where the frequencies $\omega_i$ and amplitudes $A_i$ are
initialized from the rough DFT results, and the minus sign
is consistent with the sign in Eqs.~(\ref{eq:Splus}) and (\ref{eq:Sminus}).
DFT peaks whose amplitude exceeds $5\%$ of the maximum amplitude are retained for
the fit. A global least-squares optimization is then
carried out over the full time domain $Jt \in [0,20]$.
In Fig.~\ref{fig4}a and~\ref{fig4}b, the resulting fitted time-domain curves
are overlaid on the experimental data. The optimal
fitting $\omega_i$ and amplitudes $A_i$ can be represented by
the sharp peaks in Fig.~\ref{fig4}e and~\ref{fig4}f, with an artificially set
broadening of 0.05 (larger than $100 \times$ the fitting error).

As shown in Figs.~\ref{fig4}e and \ref{fig4}f,
the even-parity channel isolates peaks at the even-sector eigenenergies $\{\tilde{E}_0^{e},\tilde{E}_1^{e}\}$,
whereas the odd-parity channel exhibits a single peak at $\{\tilde{E}_1^{o}\}$.
Comparison with the combined spectrum in Fig.~\ref{fig3}a (levels $\{0,2,3\}$)
confirms the expected $\mathbb{Z}_2$ parity assignments.
Similarly, fitting $S_{-}(t)$ for the second configuration (Fig.~\ref{fig4}c)
yields the odd-sector eigenenergy $\{\tilde{E}_0^{o}\}$ (Fig.~\ref{fig4}g),
while fitting $S_{+}(t)$ for the third configuration (Fig.~\ref{fig4}d)
gives $\{\tilde{E}_0^{e},\tilde{E}_1^{e}\}$ (Fig.~\ref{fig4}h).
In all cases, the peak distributions agree with the predicted parity classification
(see Supplementary Information Section~I).
Furthermore, parity separation cleanly resolves closely spaced levels
such as $\tilde{E}_1^{e}$ and $\tilde{E}_1^{o}$,
facilitating precise eigenenergy extraction.

Equations~(\ref{eq:Splus}) and (\ref{eq:Sminus}) show that the amplitude in each parity channel
equals the spectral weight of the corresponding eigenstate
($A_i = |c_n|^2 \equiv F_n$, where $F_n=|\langle \psi_n|\psi(\bm{\theta})\rangle|^2$
is the fidelity with respect to the $n$-th eigenstate),
so that these fidelities follow directly from the fitted amplitudes without additional measurements.
Choosing an appropriate $\tau$ maximizes the target DFT amplitude $A(\tau)$,
thereby optimizing the signal-to-noise ratio.
The fitted eigenenergies (red stars, Figs.~\ref{fig4}e--h; Table~\ref{tab1}, column~L9)
are in excellent agreement with ED results (deviations $<0.001$).

Using the scaling dimension $\Delta_{\sigma}=0.5181$, we establish the state--operator correspondence:
the parameter $R$ is fixed by the $\mathbb{Z}_2$ odd-sector splitting,
which yields the even-sector scaling dimension $\Delta_\epsilon=1.5850$,
consistent with conformal bootstrap predictions.
This gives the critical exponents $\nu=0.7067$, $\eta=0.0362$ (set by $\Delta_{\sigma}$),
together with $\beta$ and $\gamma$, all consistent with both ED and the 3D Ising CFT (Table~\ref{tab2}).
The remaining deviations are dominated by finite-size effects and diminish with increasing system size
(see Supplementary Information Section~I).

\begin{table}[t]
\caption{Eigenenergies of the two lowest-energy eigenstates in each of the $Z_2$ even and $Z_2$ odd sectors, as obtained from exact diagonalization, extended VQE simulations, and experimental results on the L9 and N13 superconducting quantum processors. The shown energies are all dimensionless with respect to the coupling strength $J$, that is, presented in the form of $\tilde{E}_i^{e,o}/J$.}
\label{tab:eigenvalues}
\renewcommand{\arraystretch}{1.0}
\setlength{\extrarowheight}{3pt}
\begin{ruledtabular}
\begin{tabular}{ccccc}
 & ED & Simulation & L9 & N13 \\
\hline
$\tilde{E}_0^{e}/J$ & $-21.1304$ & $-21.1305$ & $-21.1305$ & $-21.1305$ \\
$\tilde{E}_0^{o}/J$ & $-19.5053$ & $-19.5053$ & $-19.5057$ & $-19.5054$ \\
$\tilde{E}_1^{o}/J$ & $-16.3808$ & $-16.3810$ & $-16.3806$ & $-16.3809$ \\
$\tilde{E}_1^{e}/J$ & $-16.1593$ & $-16.1603$ & $-16.1598$ & $-16.1605$
\label{tab1}
\end{tabular}
\end{ruledtabular}
\end{table}

Although initially implemented on a topology matching the cubic Hamiltonian,
our digital protocol is inherently hardware-agnostic. To demonstrate this,
we executed the complete procedure---circuit synthesis
(Supplementary Fig.~S3b), optimization, and measurement---on an
8-qubit linear chain encoded on N13,
a 13-qubit superconducting processor with nearest-neighbor coupling. Mapping cubic couplings
onto the 1D chain via intermediate qubits yields spectral fidelities
comparable to the L9 cubic configuration (Supplementary Information
Section~III). The extracted eigenenergies
(Table~\ref{tab1}, column N13) match the L9 precision.
The primary practical difference is a restricted set of viable
$\tau$ candidates on the chain (Supplementary Fig.~S7),
which often drives the optimization toward higher eigenstates.

\begin{table}[htbp]
\centering
\caption{Scaling dimension and critical exponents obtained from the L9 and N13 superconducting quantum processors,
compared with ED and conformal bootstrap~\cite{Simmons-Duffin2017} results.}
\label{tab2}
\renewcommand{\arraystretch}{1.0}
\setlength{\extrarowheight}{3pt}
\begin{ruledtabular}
\begin{tabular}{ccccc}
 ~~~~& L9 & N13 & ED & Bootstrap \\
\hline
~~~~$\Delta_{\epsilon}$ & 1.5850 & 1.5845  & 1.5849  & 1.4126 \\
~~~~$\nu$               & 0.7067 & 0.7065  & 0.7066  & 0.6300 \\
~~~~$\beta$             & 0.3662 & 0.3660  & 0.3661  & 0.3264 \\
~~~~$\gamma$            & 1.3879 & 1.3874  & 1.3877  & 1.2371 \\
\end{tabular}
\end{ruledtabular}
\end{table}

\medskip

\noindent\bluefour{\it Discussions.}---
Our work demonstrates that
the critical properties of the 3D Ising universality class can be
reliably extracted using a few accurately determined low-lying
eigenenergies, obtained via a hybrid quantum-classical protocol
on a programmable superconducting processor. The key to this extraction
is a phase-based loss function within an extended VQE framework,
which provides a reliable access to excited states without introducing
explicit orthogonality constraints or subspace projections. The protocol
is inherently scalable. The number of variational parameters grows
linearly with the number of qubits, and the spectral analysis depends
only on the quality of the prepared state, not on the system's
Hilbert-space dimension.
In the current experiment, we extract the four lowest energy levels
of the TFIM on the cube lattice.
The results, summarized in Table~\ref{tab:eigenvalues}, are in an excellent
agreement with both the classical simulation and the exact diagonalization.
Following the theoretical framework, we use the scaling dimension $\Delta_{\sigma}$ as an
input to establish the state-operator correspondence. In this way, we obtain the scaling dimension
$\Delta_{\epsilon}$ and the 3D Ising critical exponents, as shown in Table~\ref{tab2}.

For the system sizes simulated here, the quantum simulation results
agree well with exact diagonalization, while deviations from the
theoretical critical exponents are dominated by finite-size effects.
As the number of qubits increases, the extracted critical exponents
will become more accurate; the corresponding results from 6 to 20 qubits
are presented in Supplementary Information Section I. In addition,
larger system sizes allow more low-lying energy levels to be used to
characterize the critical-point data. The critical point and the
state-operator correspondence can then be determined directly from the spectrum, thus all critical
exponents can in principle be extracted without using $\Delta_{\sigma}$
as an input. Beyond the extraction of critical exponents, a more
ambitious experimental objective is to access the CFT spectrum, which
may become achievable as the number of qubits increases.
Strategies for increasing the number of qubits are discussed in the Methods.



More broadly, this strategy can in principle be generalized to other two-dimensional or 3D
CFTs. A particularly interesting target is the two-dimensional Yang--Lee CFT.
As an unconventional critical theory, it provides a valuable
setting in which the direct determination of the universal critical data remains
experimentally scarce.~\cite{PhysRevLett.131.080403} Meanwhile, in two
dimensions the low-energy spectrum is generally easier to resolve and to
match with the conformal data than in three dimensions, making the high-precision
extraction more feasible. This suggests that the same approach may allow
the critical exponents of the Yang--Lee theory to be determined with high
accuracy from only about ten qubits. While these prospects are encouraging,
the quantum simulation of the 3D CFT spectrum remains a formidable
challenge. In this sense, the present work is a first step toward the
longer-term goal of experimentally probing the universal conformal data in
three dimensions.


\noindent\emph{Note added:} Near the completion of this work, we become
aware of Ref.\onlinecite{wu2026qubitdiscretizationsd3conformal}, which also makes
the theoretical proposal that quantum simulation on Platonic lattices
can provide access to 3D conformal data. In this paper,
our simulation of critical exponents and the implementation from theory can be viewed as the first step toward this
more general goal.


%
\makeatletter
\newcommand{\sectionNoTOC}[1]{%
  \par
  \addvspace{1.2\baselineskip}
  \penalty-300
  \noindent
  \begingroup
    \centering\bfseries #1\par
  \endgroup
  \nopagebreak
  \vspace{0.6\baselineskip}
  \@afterheading
}
\newcommand{\methodsubsec}[1]{%
  \par
  \addvspace{0.9\baselineskip}
  \penalty-300
  \noindent{\bfseries #1}\par
  \nopagebreak
  \vspace{0.35\baselineskip}
  \@afterheading
}
\makeatother

\sectionNoTOC{Data Availability}
The datasets generated and analysed during the current study are
available via Zenodo at
https://doi.org/10.5281/zenodo.20506684.

\sectionNoTOC{Code Availability}
The code used to analyse the datasets during the current study is available via Zenodo
at https://doi.org/10.5281/zenodo.20569378.

\sectionNoTOC{Acknowledgments}
H.S. and G.v.C. are by Quantum Science and Technology-National
Science and Technology Major Project with grant No.~2025ZD0300500,
and by NSFC with Grants No.~92565110
and No.~12574061.
The authors from Zhejiang University (Y.Y. and colleagues) thank the
support from the National Natural Science Foundation of China (Grant No. 12074336),
the National Key Research and Development Program of
China (Grant No. 2025YFH0102104, and No. 2022YFA1403202). Y.Y. also acknowledges the funding
support from Tencent Corporation.

\sectionNoTOC{Author contributions}
L.S. and Y.Y. designed the experimental protocol. L.S. performed the experiment. H.S. and G.v.C. developed the theory.
S.H. designed the superconducting processor. K.B. fabricated the processor. Y.L. and S.H. calibrated the processor.
X.Y. designed and fabricated the JPA.
Z.Z. helped with the algorithm of error correction. X.L., W.J., B.C., X.Z., and J.W. assisted with the protocol and numerical simulations.
Y.Z., S.Z, G.v.C., and Y.Y. supervised the project. All authors contributed to discussing the results
and writing the manuscript.

\sectionNoTOC{Competing interests}
The authors declare no competing interests.



\makeatletter
\begingroup
\let\addcontentsline\@gobblethree

\endgroup
\makeatother

\clearpage

\makeatletter
{\let\addcontentsline\@gobblethree
\section{Methods}
}
\makeatother

\methodsubsec{Variational quantum circuit architecture}

We employ a two-layer, hardware-efficient variational circuit comprising
$N_p = 16$ independent parameters (one per qubit in each rotation layer).
The unitary evolution is given by
\begin{equation}
U(\bm{\theta}) = \mathcal{E}_2\mathcal{R}_2(\bm{\theta}_{9:16})\,
\mathcal{E}_1\mathcal{R}_1(\bm{\theta}_{1:8}),
\label{eq:ansatz}
\end{equation}
where $\mathcal{R}_l$ denotes a layer of single-qubit $R_y$ rotations
and $\mathcal{E}_l$ represents an entangling layer of controlled-NOT
(CNOT) gates. The connectivity of the entangling blocks is tailored
to the native hardware topology, mapping the twelve edges of a cube.
Specifically, $\mathcal{E}_1$ applies CNOT gates along these edges
in a predefined sequence, whereas $\mathcal{E}_2$ reverses this order,
making their entanglement patterns exact structural inverses.
In the experiment, each CNOT gate is compiled into native controlled-$Z$
(CZ) gates and single-qubit $R_y$ rotations (Supplementary Fig.~S3c).

\methodsubsec{Optimization protocol and gradient evaluation}

The variational parameters are optimized using the limited-memory
Broyden--Fletcher--Goldfarb--Shanno with box constraints
(L-BFGS-B) algorithm~\cite{Nocedal2016book},
a gradient-based quasi-Newton method. At each iteration $i$,
the exact gradient of the loss function is evaluated analytically
via the parameter-shift rule~\cite{MitaraiPRA18}. For each parameter
$\theta_l \in \bm{\theta}^{(i-1)}$, we define the unshifted state
$|\psi_l\rangle = U({\boldsymbol{\theta}})|\psi_0\rangle$ and the shifted states
$|\psi^\pm_l\rangle = U({\boldsymbol{\theta}}^\pm_l)|\psi_0\rangle$,
where ${\boldsymbol{\theta}}^\pm_l = \{\dots, \theta_l\pm\pi/2, \dots\}$.
The partial derivative is then computed as
\begin{equation}
\frac{\partial \mathcal L_i}{\partial \theta_l}
=
-\frac{1}{2|\mathcal W|}
\mathrm{Re}\left[\left(\mathcal W^+_l-\mathcal W^-_l\right)\mathcal W^\ast\right],
\label{eq:derivative}
\end{equation}
where $\mathcal W_l = \langle\psi_l|e^{-iH_{\mathrm{p}}\tau}|\psi_l\rangle$ and
$\mathcal W^\pm_l = \langle\psi^\pm_l|e^{-iH_{\mathrm{p}}\tau}|\psi^\pm_l\rangle$.
These gradients inform the L-BFGS-B update step, yielding the optimized parameter
set $\bm{\theta}^{(i)}$ and the corresponding state
$\ket{\psi(\bm{\theta}^{(i)})} = U(\bm{\theta}^{(i)})\ket{\psi_0}$ for the next iteration.

\methodsubsec{State characterization and error mitigation}

At each optimization step, the 8-qubit variational state is prepared
and measured on the L9 superconducting processor. Because the target
Hamiltonian is real and symmetric, its eigenstates can be strictly
represented by real amplitudes in the computational basis. We exploit
this property by employing an $R_y$-only ansatz, which restricts
the evolution to the real subspace and ensures that all variational
states remain real superpositions. Consequently, we can reconstruct
the wavefunction efficiently without full quantum state tomography
(QST). We sample the 256-outcome probability distribution in the
computational basis using $N_{\mathrm{shots}} = 5000$ measurements,
obtaining the raw probability vector $\bm{p}_\mathrm{raw}$.
To mitigate measurement inaccuracies, we apply a pre-calibrated
$256 \times 256$ readout-error correction matrix $M$, yielding the
corrected distribution $\bm{p}_\mathrm{corr}=M^{-1}\bm{p}_\mathrm{raw}$.
The experimental state vector is then constructed by assigning the sign
of each amplitude based on a numerical reference state:
$\ket{\psi_\mathrm{exp}}=\sum_S \mathrm{sgn}(\psi_{\mathrm{ref},S})\sqrt{p_{\mathrm{corr},S}}\ket{S}$.

\methodsubsec{Strategies for increasing the number of qubits}

Increasing the number of qubits is essential for both improving the accuracy
and extracting more scaling dimensions. Besides the experimental
challenge of preparing and controlling larger numbers of qubits, two
further issues deserve attention. The first is that the Hilbert space
grows exponentially with the number of qubits, making it increasingly
difficult to accurately resolve the few lowest energy levels from the
much larger many-body spectrum. Although a single-shot state measurement
has been applied in current experiment, the optimization loop in general
relies on the repeated quantum-state measurements to compute the gradients---a
procedure whose sampling overhead grows with system size. Future
implementations could employ advanced measurement strategies, such as
the shadow tomography, to access the quantum state in a single-shot or
few-shot manner~\cite{TanDianPRL2024,YouZhouPRAppl2025}.
The second issue is how to increase the number of qubits beyond the
twenty qubits limit of the Platonic lattices while preserving a spherical
geometry. One natural direction is to study the model on Archimedean
lattices that retain the $I_h$ symmetry, while another is to refine the
spherical discretization by subdividing the triangular faces of an
icosahedron into smaller triangles~\cite{wu2026qubitdiscretizationsd3conformal}.

\clearpage

\makeatletter
\global\let\@FMN@list\@empty
\let\present@bibnote\@gobbletwo
\let\present@FM@footnote\@gobbletwo
\makeatother

\widetext

\begin{center}
\textbf{\large Supplementary Information for  ``3D Ising criticality with Platonic lattice superconducting qubits''}
\end{center}

\addtocontents{toc}{\protect\setcounter{tocdepth}{1}}
{
\tableofcontents
}

\renewcommand{\thefigure}{S\arabic{figure}}
\setcounter{figure}{0}
\renewcommand{\theequation}{S\arabic{equation}}
\setcounter{equation}{0}
\renewcommand{\thesection}{\Roman{section}}
\setcounter{section}{0}
\setcounter{secnumdepth}{4}

\section{Exact Diagonalization of the Transverse Field Ising Model on Platonic Lattices}

\begin{figure}[htbp]
    \centering
    \includegraphics[width=0.55\linewidth]{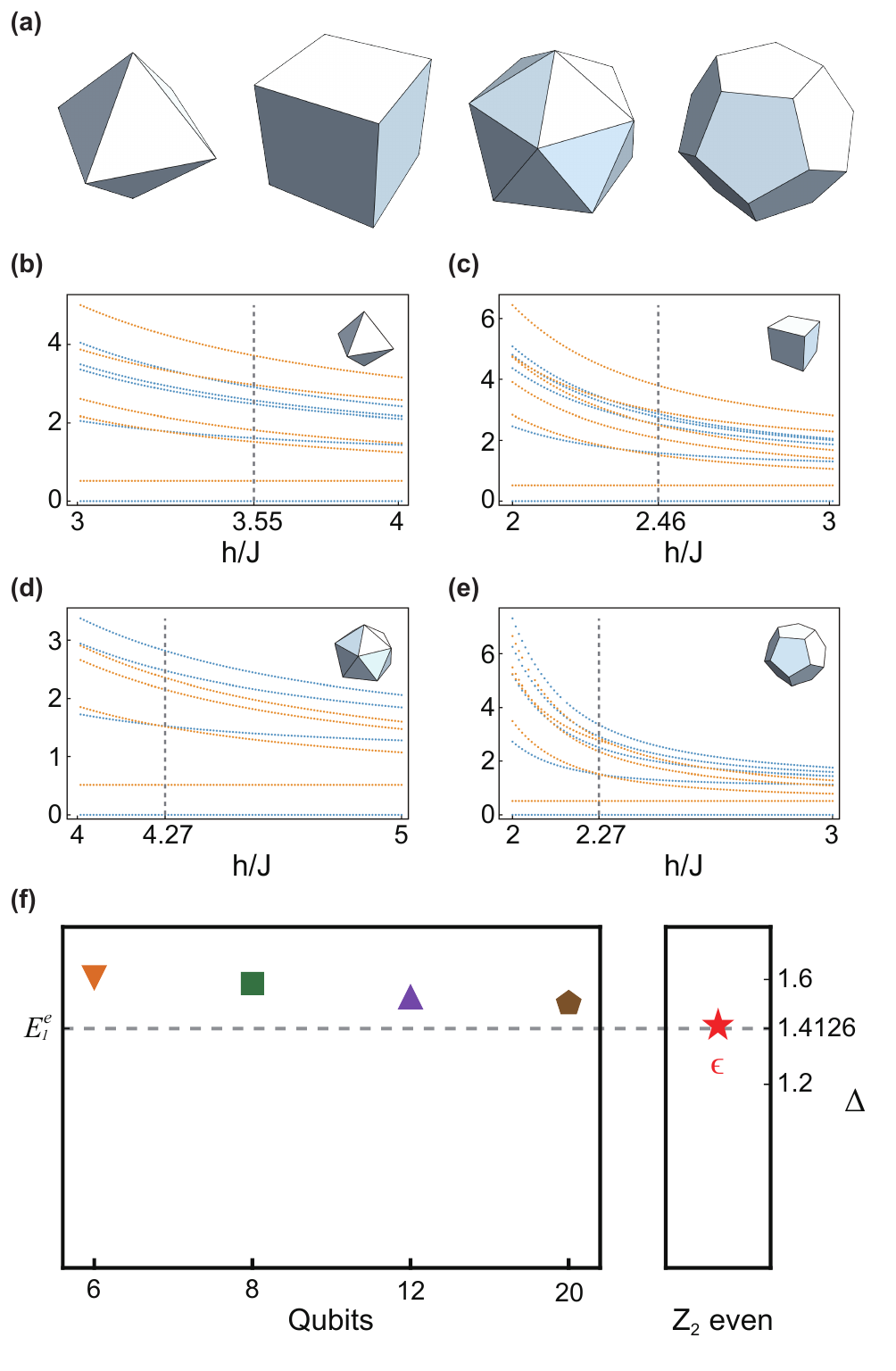}
    \caption{Exact diagonalization results for the TFIM on Platonic lattices. (a) The octahedron, cube, icosahedron and dodecahedron lattices, corresponding to 6, 8, 12, and 20 qubits. (b)-(e) Low-lying renormalized energy levels for octahedron, cube, icosahedron and dodecahedron lattices with different transverse field $h$. The blue points denote the levels in the $\mathbb{Z}_2$ even sector, while the orange points denote the levels in the $\mathbb{Z}_2$ odd sector. The critical field $h_c$ is determined by the condition $E_1^o-E_0^o=1$, which is indicated by the gray dashed line. (f) Summary of the extracted $E_1^e$ values for different lattices at the critical point, with the red marker denoting the reference scaling dimension of the 3D Ising CFT.}
    \label{figS1}
\end{figure}

In this section, we present the details of the exact diagonalization of the transverse field Ising model (TFIM) on Platonic lattices.
The Platonic lattices considered in this work are the octahedron, cube, icosahedron, and dodecahedron, as shown in Fig.~\ref{figS1}(a). The qubits are placed on the vertices of these lattices, which correspond to the sites of the TFIM, while the edges represent the interactions.
The TFIM on a Platonic lattice is

\begin{equation}
H_{\mathrm{p}} = - J \sum_{\langle ij\rangle} \sigma_i^z \sigma_j^z - h \sum_i \sigma_i^x ,
\end{equation}
where $\sigma_i^{x,y,z}$ are Pauli matrices, and $\langle ij\rangle$ denotes nearest-neighbor pairs on the Platonic lattice.
The parameter $J$ denotes the Ising coupling, and $h$ denotes the transverse-field strength.

The Hamiltonian is invariant under the global $\mathbb{Z}_2$ symmetry generated by $U=\prod_i \sigma_i^x$, which classifies the eigenstates into $\mathbb{Z}_2$ even and $\mathbb{Z}_2$ odd sectors.

For each value of the transverse field $h$, we obtain the eigenvalues $\tilde{E}_i^{e,o}$ by performing exact diagonalization (ED). If the system is tuned to criticality, $h = h_c$, the low-lying energy levels $\tilde{E}_i^{e,o} - \tilde{E}_0^{e}$ are expected to be proportional to the scaling dimensions of the 3D Ising CFT. We fix the proportionality constant with the theoretical scaling dimension $\Delta_{\sigma}$, which gives
\begin{equation}
    R = \frac{\Delta_{\sigma}}{\tilde{E}_0^{o} - \tilde{E}_0^{e}}.
\end{equation}
The renormalized energy levels are then defined as
\begin{equation}
    E_i^{e,o} = (\tilde{E}_i^{e,o} - \tilde{E}_0^{e}) R.
\end{equation}
The renormalized energy levels for different lattices are presented in Fig.~\ref{figS1}(b)-(e), with blue points denoting the levels in the $\mathbb{Z}_2$ even sector and orange points denoting the levels in the $\mathbb{Z}_2$ odd sector.
In the continuous limit, these renormalized energy levels at critical point are exactly equal to the scaling dimensions of the CFT. For the Platonic lattices considered here, although the systems contain only a finite number of sites, the lowest few renormalized energy levels are still expected to provide good approximations to the corresponding scaling dimensions.

To determine the critical field $h_c$, we consider the two lowest renormalized energy levels in the $\mathbb{Z}_2$ odd sector. In the CFT description, these two levels correspond to the primary operator $\sigma$ and its descendant $\partial_{\mu}\sigma$, whose scaling dimensions differ by 1 exactly. We therefore determine the critical field $h_c$ by imposing
\begin{equation}
    E_1^o-E_0^o=1,
\end{equation}
where $E_0^o$ and $E_1^o$ are the first and second renormalized energy levels in the $\mathbb{Z}_2$ odd sector. For each Platonic lattice, the critical transverse field $h_c$ obtained from this criterion is marked by the gray dashed line in Fig.~\ref{figS1}(b)--(e).

With $h_c$ determined, we turn to the $\mathbb{Z}_2$ even sector. The lowest level $E_0^e$ is zero, while the second level $E_1^e$ corresponds to the scaling dimension $\Delta_{\epsilon}$. The values of $E_1^e$ at $h_c$ extracted for different Platonic lattices are summarized in Fig.~\ref{figS1}(f).
As shown in the figure, improving the lattice symmetry from $O_h$ to $I_h$ has a much more pronounced effect on the extracted results than simply increasing the number of qubits. This is why estimates with acceptable accuracy can be obtained using only a modest number of qubits. Another reason lies in the structure of the Ising CFT spectrum itself, the lowest few energy levels are well separated from the higher ones, making their identification more robust in finite size simulations.

We can further determine the 3D Ising critical exponents $\nu$, $\beta$, and $\gamma$ through
\begin{equation}
\nu = \frac{1}{d-\Delta_{\epsilon}},
\end{equation}
\begin{equation}
\beta = \frac{\nu}{2}(d-2+\eta),
\end{equation}
and
\begin{equation}
\gamma = \nu(2-\eta).
\end{equation}
Since $\Delta_{\sigma}$ has already been used as an input to fix the proportionality constant in the spectrum renormalization, $\eta$ is not extracted  from the finite size spectrum. Here, we take the theoretical value of $\eta$ and combine it with the extracted $\nu$ to determine the other critical exponents. The results are summarized in Table~\ref{tab3}.

\begin{table}[htbp]
\centering
\renewcommand{\arraystretch}{1.3}

\begin{tabular}{l @{\hspace{10pt}} c @{\hspace{30pt}} c @{\hspace{30pt}} c @{\hspace{20pt}} c @{\hspace{20pt}} c}
\hline\hline
~~~~~~~~ & Octahedron & Cube & Icosahedron & Dodecahedron &  Theory\\
\hline
~~~~$\Delta_{\epsilon}$ & 1.6143 & 1.5849 & 1.5261 & 1.5093 & 1.4126 \\
~~~~$\nu$               & 0.7216 & 0.7066 & 0.6785 & 0.6708 & 0.6300 \\
~~~~$\beta$             & 0.3739 & 0.3661 & 0.3515 & 0.3475 & 0.3264 \\
~~~~$\gamma$            & 1.4172 & 1.3877 & 1.3324 & 1.3173 & 1.2371 \\
\hline\hline
\end{tabular}
\caption{Critical exponents extracted from different Platonic lattices, compared with the theoretical values of the 3D Ising CFT.}
\label{tab3}
\end{table}

\section{Experimental Setup and Circuit Mapping of the L9 Chip}

\begin{figure}[htbp]
    \centering
    \includegraphics[width=1.0\linewidth]{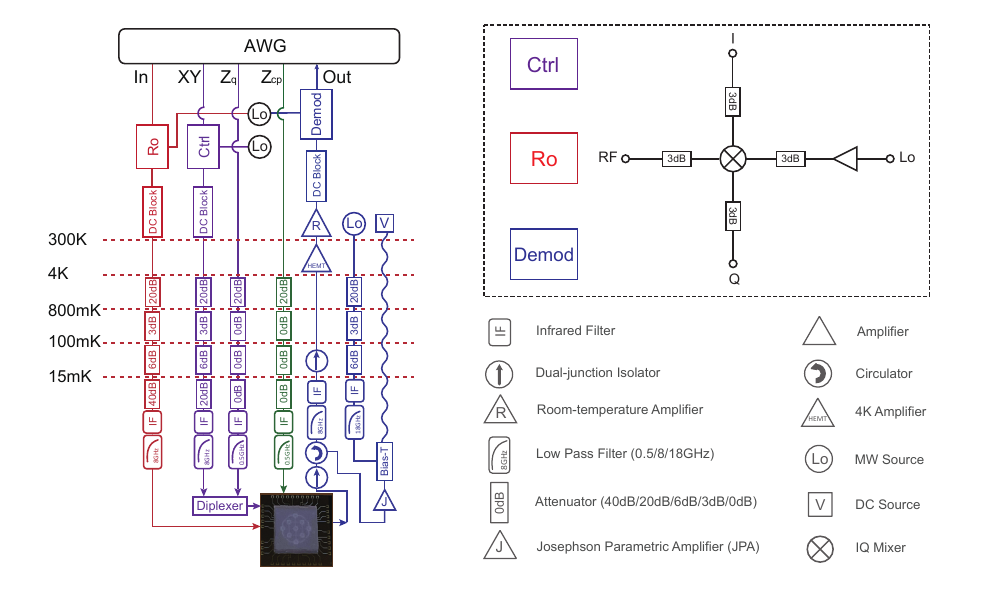}
    \caption{The experimental setup. It includes a schematic of the signal routing from room-temperature electronics to the cryogenic quantum chip. Device annotations---covering both room-temperature instrumentation (\textit{e.g.}, AWG) and cryogenic components (\textit{e.g.}, attenuators, isolators, circulators, HEMT)---are provided in the lower-right corner. In the upper-right corner, a dedicated inset details the mixer-based upconversion and downconversion architecture used for qubit control and readout.}
    \label{figS2}
\end{figure}

The experiments in the main text were performed on the 9-qubit
L9 superconducting processor. A schematic of the measurement
setup is shown in Supplementary Fig.~\ref{figS2}. The L9 chip was
wire-bonded in a commercial aluminum sample box (44 connectors)
and mounted in a standard dilution refrigerator with a
base temperature around 15 mK. The bottom square box presents
an optical micrograph of the flip-chip assembly: the upper die
contains the superconducting qubit array, and the lower die
integrates the coplanar waveguide resonators along with their
control and readout wiring. Owing to the semi-transparency of
the sapphire substrate, the spatial arrangement and interconnect
topology of the nine qubits remain discernible through the upper die.
The wiring comprises separate
lines for qubit $XY$- and $Z$-control (diplexer combined at the 15 mK stage),
coupler $Z$-control, three groups of readout input/output lines,
and pump lines for the Josephson parametric amplifiers (JPA).
Attenuators are installed on the control and input coaxial
lines at different temperature stages. On the 15 mK flange,
a commercial $K\&L$ low-pass filter and a home-built Eccosorb infrared
filter are inserted before the lines enter the sample box, which
itself is enclosed in concentric aluminum and cryoperm magnetic shields.

Qubit $XY$-control pulses are generated by a multi-channel
arbitrary-waveform generator (AWG), up-converted with an IQ mixer
using a microwave local oscillator (LO), and delivered to the device.
$Z$-control signals for the qubits and tunable couplers (including
static $DC$ offsets) are similarly supplied by AWG channels.
Readout pulses are generated by an AWG, up-converted, and sent to
the device; each readout line with multiplexed signals transmitted
through three are amplified by a JPA, followed by
a high-electron-mobility-transistor (HEMT) amplifier at 4 K and
a room-temperature amplifier. The output is then down-converted
with another IQ mixer and digitized by an analog-to-digital
converter (ADC) board for I/Q demodulation.

\begin{table}[htbp]
\centering
\renewcommand{\arraystretch}{1.3}

\begin{tabular}{lccccc}
\hline\hline
 ~~~~Qubit     & ~~~$T_1(\mu s)$~~    & ~~$T_2^{\mathrm{echo}}(\mu s)~~   $    &~~~SqRB~~~ &~ MeasFid $\ket{0}$&~~MeasFid $\ket{1}$~~~~ \\
\hline
  ~~~~~~$Q_1$            & 57.909 & 47.294 & 99.947\% & 99.246\% & 93.005\%   \\
  ~~~~~~$Q_2$            & 37.919 & 29.368 & 99.938\% & 99.257\% & 97.871\%   \\
  ~~~~~~$Q_3$            & 65.749 & 47.192 & 99.906\% & 99.568\% & 96.304\%   \\
  ~~~~~~$Q_4$            & 85.647 & 13.814 & 99.924\% & 99.280\% & 97.203\%   \\
  ~~~~~~$Q_5$            & 50.598 & 12.057 & 99.922\% & 98.890\% & 97.234\%   \\
  ~~~~~~$Q_6$            & 56.000 & 17.000 & 99.915\% & 98.974\% & 95.602\%   \\
  ~~~~~~$Q_7$            & 45.398 & 10.271 & 99.925\% & 99.400\% & 98.357\%   \\
  ~~~~~~$Q_8$            & 50.095 & 37.243 & 99.928\% & 99.632\% & 97.674\%   \\
  \hline
  ~~~~~Avg              & 57.289 & 26.779 & 99.926\% & 99.281\% & 96.656\%   \\
\hline\hline
\end{tabular}
\caption{Single-qubit coherence times, randomized benchmarking (RB) gate fidelities, and measurement fidelities.}
\label{tab4}
\end{table}

The L9 chip was fully measured, calibrated and optimized at low temperatures
for subsequent experiments.
In the main text, Fig. 1(c) illustrates the native connectivity of
the L9 processor. The architecture features a central qubit ($Q_1$)
coupled to all eight peripheral qubits ($Q_2$--$Q_9$); the peripheral
qubits are further interconnected through
sixteen tunable couplers. This high-connectivity graph natively
embeds the 12-edge topology of a cubic lattice. For the cubic Platonic lattice simulation,
we use the eight-qubit subset $Q_1$--$Q_8$. The mapping from the
logical circuit qubits $\{q_1,\dots,q_8\}$ to these physical
qubits has been given in Fig. 1(d).

Table~\ref{tab4} summarizes key coherence and operational parameters
of the 8 superconducting qubits on the L9 chip. The average energy-relaxation time
$T_1$ is $57.289~\mu\text{s}$---a notably high value despite the
presence of multi-connected qubit-qubit interactions.
Similarly, the average spin-echo dephasing time $T_2^\text{echo}$ is $26.779~\mu\text{s}$,
confirming that these interactions induce only minimal additional phase
decoherence. During characterization, all tunable couplers were held in
their off state (zero coupling), and each qubit was biased at its individually
optimized cooperative operating point---selected to balance anharmonicity,
coherence, and gate performance for subsequent quantum circuit execution.
State discrimination fidelities for $\ket{0}$ and $\ket{1}$ are reported per qubit
in Table~\ref{tab4}. Under these conditions, all qubits achieve single-qubit
gate fidelities exceeding 99.9\%, as verified by interleaved randomized
benchmarking (see Table~\ref{tab4}).

\begin{table}[htbp]
\centering
\renewcommand{\arraystretch}{1.3}

\begin{tabular}{cccc}
\hline\hline
~~~~Edge & ~~~~Vertex pair $(Q_i,Q_j)$~~~~ & ~~~~Vertex pair $(q_i,q_j)$~~~~ &~~~~ CZ gate RB Fid~~~~~~~~ \\
\hline
~~~~1 & (1,2) & (5,4) & 99.767\% \\
~~~~2 & (1,6) & (5,6) & 98.919\% \\
~~~~3 & (1,8) & (5,8) & 98.357\% \\
~~~~4 & (2,3) & (4,3) & 99.547\% \\
~~~~5 & (2,5) & (4,1) & 99.531\% \\
~~~~6 & (3,4) & (3,2) & 99.940\% \\
~~~~7 & (3,6) & (3,6) & 99.556\% \\
~~~~8 & (4,5) & (2,1) & 99.773\% \\
~~~~9 & (4,7) & (2,7) & 99.787\% \\
~~~~10 & (5,8) &(1,8) & 99.534\% \\
~~~~11 & (6,7) & (6,7) & 99.260\% \\
~~~~12 & (7,8) & (7,8) & 99.521\% \\
\hline
~~~~AVG&  & &99.458\%\\
\hline\hline
\end{tabular}
\caption{Two-qubit CZ gate fidelity measured by randomized benchmarking on all edges of the cubic lattice.}
\label{tab5}
\end{table}

For the 12-edge topology of the cubic lattice, Table~\ref{tab5} summarizes
the 12 edge-qubit assignments for the cubic lattice topology: for each edge
$e_k$ ($k = 1,\dots,12$), it lists the physical qubit pair $(Q_i, Q_j)$ and the
corresponding cuicuit qubit indices $(q_i, q_j)$. Also reported are the two-qubit
CZ gate fidelities, extracted via interleaved randomized benchmarking.
All but two gates---those associated with edges 2 and 3---achieve fidelities
above 99.0\%, with nine of the twelve exceeding 99.5\%. These results
demonstrate high-fidelity entangling operations across a densely
interconnected architecture, representing a significant achievement
for the multi-qubit quantum processor L9.

\begin{figure}[htbp]
    \centering
    \includegraphics[width=1.0\linewidth]{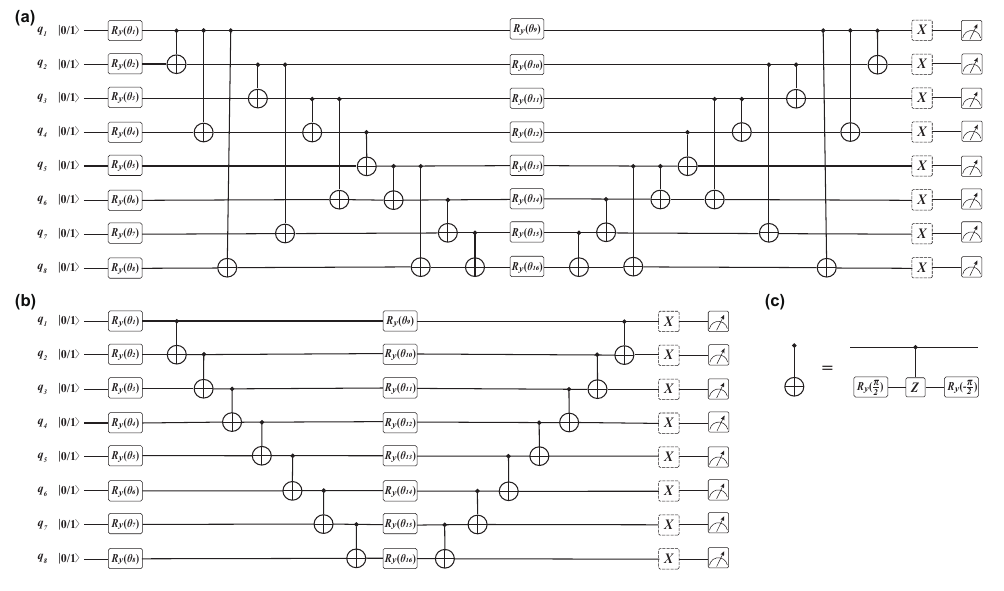}
    \caption{Quantum circuit diagram. (a) State-preparation circuit implemented on the eight selected qubits of the L9 chip. (b) State-preparation circuit implemented on the linear 8-qubit chain of the N13 chip. (c) Native-gate decomposition of the CNOT gate as implemented on the hardware.}
    \label{figS3}
\end{figure}

Correspondingly, the Supplementary Fig.~\ref{figS3}(a)
shows the compiled two-layer hardware-efficient ansatz used in
the experiment: each layer alternates parallel single-qubit $R_Y$
rotations with a fixed pattern of CNOT gates that follows the
12 cubic edges $e_k$. The CNOT gates are decomposed into the processor's
native gates (single-qubit rotations and CZ gates) according
to this physical mapping (Fig.~\ref{figS3}(c)).

\begin{figure}[htbp]
    \centering
    \includegraphics[width=0.95\linewidth]{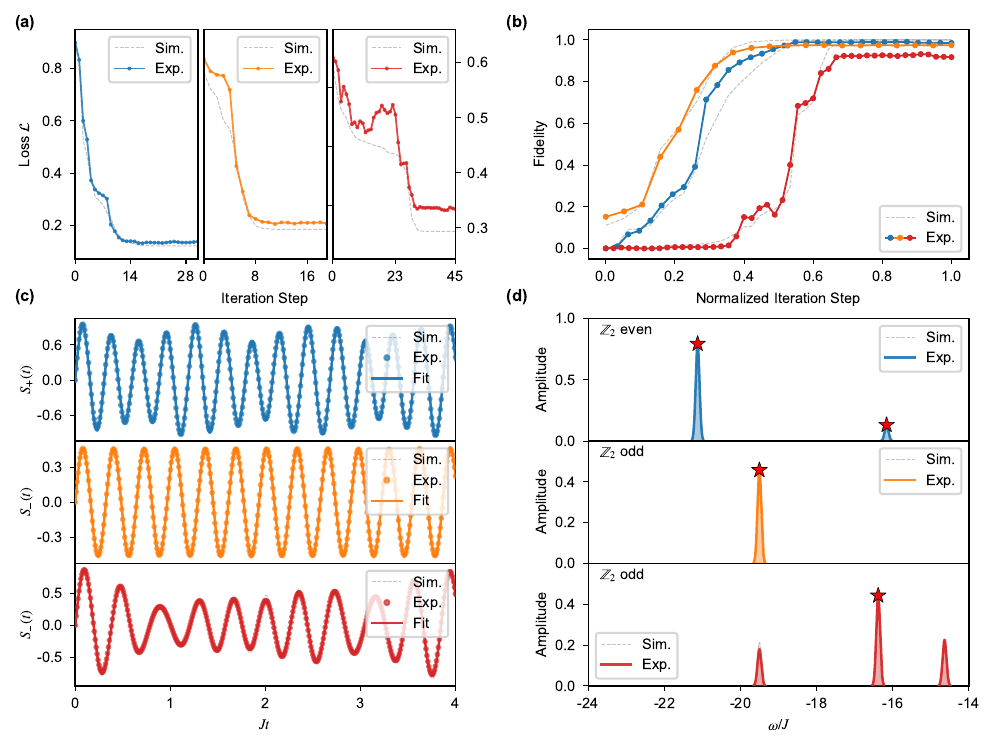}
    \caption{Variational optimization of quantum states
    for three initial configurations on an 8-qubit linear chain in a 13-qubit processor.
    (a) Loss function $\mathcal{L}$ versus optimization iteration step for
    initial states $\ket{00000101}$, $\ket{00000010}$,
    and $\ket{00000000}$, with corresponding evolution times
    $J\tau = 17.4$, $12.7$, $6.4$, respectively. Grey dashed curves denote numerical
    simulation; solid circles connected by lines show experimental data (blue,
    orange, red).
    (b) State fidelity with respect to the final converged state as a
    function of normalized iteration step. The same color scheme and line styles as in (a) are used.
    (c) Imaginary part of the parity-resolved time-domain signals for the three initial configurations.
    Top: even-channel signal $S_+(t)$ for the state prepared from $\ket{00000101}$ ($J\tau = 17.4$).
    Middle: odd-channel signal $S_-(t)$ for the state from $\ket{00000010}$ ($J\tau = 12.7$).
    Bottom: odd-channel signal $S_-(t)$ for the state from $\ket{00000000}$ ($J\tau = 6.4$).
    Grey dashed curves show numerical simulations;
    solid colored circles represent experimental data; solid colored lines are fits to the data.
    (d) Amplitude spectra obtained from multi-component sinusoidal fitting of the signals in (c), plotted versus angular frequency $\omega/J$.
    Top: spectrum of the even-channel signal in (c), showing peaks at eigenenergies $\{\tilde{E}_0^{e},\tilde{E}_1^{e}\}$.
    Middle: spectrum of the middle odd-channel signal in (c), retaining eigenenergy $\{\tilde{E}_0^{o}\}$.
    Bottom: spectrum of the bottom odd-channel signal in (c), retaining eigenenergies $\{\tilde{E}_0^{o},\tilde{E}_1^{o},\tilde{E}_2^{o}\}$.
    All fitted peaks have an artificially set Gaussian broadening of 0.05, exceeding $100\times$ the fitting error.
    Red stars mark the extracted eigenenergies and the corresponding state fidelities (amplitudes) for the four lowest eigenenergy levels.
   }
    \label{figS4}
\end{figure}

\section{13-qubit Processor and Experimental results on N13 Chip}

\begin{figure}[htbp]
    \centering
    \includegraphics[width=0.55\linewidth]{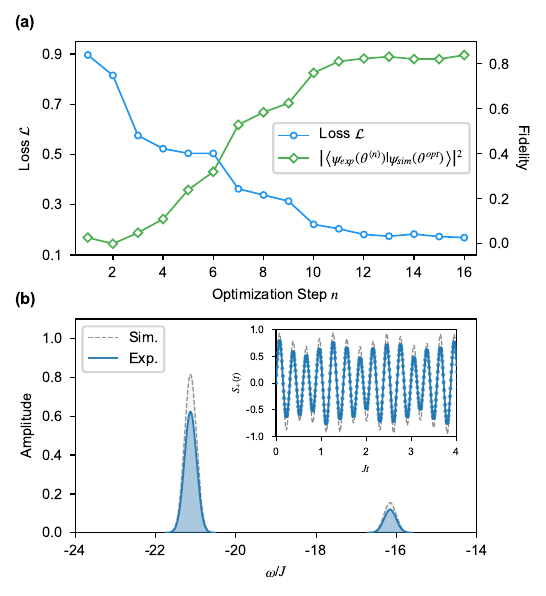}
    \caption{An optimization procedure employing gradients estimated directly
    from experimental measurements. (a) Loss function and state fidelity versus
    optimization step. Fidelity is evaluated with respect to the final simulated
    state. (b) Main panel: amplitude spectrum of the final prepared state,
    showing peaks at eigenenergies $\{\tilde{E}_0^{e},\tilde{E}_1^{e}\}$. Inset: the corresponding even-channel
    time-domain signal $S_+(t)$ for the state prepared from $\ket{00000101}$ ($J\tau = 17.4$).
    }
    \label{figS5}
\end{figure}

Optimal algorithmic implementation typically benefits from aligning the
qubit connectivity graph with the interaction topology of the target
Hamiltonian. However, our inherently digital approach imposes no strict
constraints on this mapping. To further validate this flexibility,
we performed a cross-validation experiment on a different 13-qubit processor (N13). We selected eight physical qubits in a
linear chain on the device and implemented the two-layer
hardware-efficient ansatz shown in Fig.~\ref{figS3}(b),
following a variational optimization procedure analogous
to that described in the main text.

As shown in Supplementary Fig.~\ref{figS4}(a), the loss function versus
iteration step is plotted for the three initial states on the N13
linear chain. Supplementary Fig.~\ref{figS4}(b) presents the corresponding
optimization fidelities as a function of the normalized iteration step.
The convergence behavior and final fidelities (reaching $\sim$92--98\%)
are comparable to those obtained on the cube-connected L9 processor,
further confirming the algorithmic robustness to hardware connectivity.

Supplementary Fig.~\ref{figS4}(c) displays the parity-resolved time-domain
signals $S_+(t)$ or $S_-(t)$ for the three configurations measured
on the linear chain, and Supplementary Fig.~\ref{figS4}(d) presents
the corresponding amplitude spectra obtained from multi-component
sinusoidal fitting. The spectral peaks clearly identify the four lowest
eigenenergies $\{\tilde{E}_0^{e},\tilde{E}_0^{o},\tilde{E}_1^{o}\,\tilde{E}_1^{e}\}$, consistent with the main experiment.
The extracted eigenenergies ($\omega/J$, marked by red stars) agree with
theoretical values to within 0.001, validating the accuracy of the
spectral estimation protocol on a distinctly connected hardware graph.

Finally, Supplementary Fig.~\ref{figS5}(a) presents a complete optimization
loop driven by gradients obtained directly from experimental measurements
via the parameter-shift rule, rather than from classical simulation.
The plot shows the fidelity of the prepared state relative to the
simulated final state as a function of the iteration step.
Starting from an initial parameter set, the optimization converges
to a fidelity above 83\% within 15 steps. While the achieved fidelity
is slightly lower, the resulting deviation in eigenstate distribution
does not affect the construction of the time-domain signal,
its Fourier transform, and the subsequent fine fitting to extract
eigenenergies, as confirmed in Supplementary Fig.~\ref{figS5}(b).
This outcome confirms the feasibility of using in-situ experimental
gradients to steer the variational search---a key step toward fully
autonomous calibration of the algorithm on larger-scale quantum
devices where classical simulation becomes impractical.

\section{Selection of the Evolution Time \texorpdfstring{$\tau$}{tau} Parameter in Phase-based Loss Function}

\begin{figure}[htbp]
    \centering
    \includegraphics[width=0.9\linewidth]{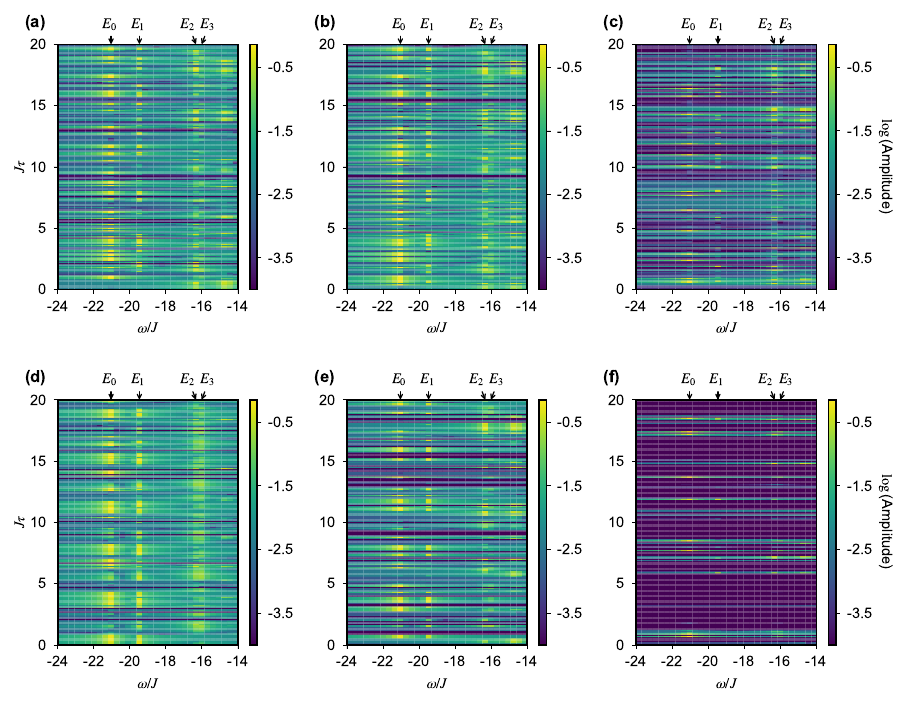}
    \caption{Fourier transform spectra for different initial states and $\tau$ parameters in loss function.
    All spectra are obtained from numerical simulations. (a--c) Results for the 8-qubit
    cube connectivity on the L9 processor, using the final state prepared by the
    variational circuit in Fig.~\ref{figS3}(a). From top to bottom: spectra for
    initial states $|00000001\rangle$, $|00000010\rangle$, and $|00000101\rangle$.
    Horizontal axis shows normalized frequency $\omega/J$; vertical axis
    shows the evolution time $\tau$ (normalized as $J\tau$ ). Color represents the logarithm of
    the Fourier amplitude. Bright discontinuous vertical lines indicate
    the positions of the four lowest eigenenergies $E_0$--$E_3$. For each eigenenergy, a
    $\tau$ value corresponding to a large amplitude can be selected for fine fitting
    and accurate energy extraction. (d--f) Corresponding simulated spectra for the same
    three initial states implemented on the 8-qubit linear chain with the circuit
    in Fig.~\ref{figS3}(b).}
    \label{figS6}
\end{figure}
As described in the main text, optimizing a chosen initial state with
a fixed evolution time $\tau$ under the phase-based loss function
yields a final state that is a superposition of a few eigenstates.
A coarse scan of $\tau$ provides initial estimates of the eigenenergies
without prior knowledge. For different trial configurations ($\ket{\psi_0},\tau$),
the initial state is optimized to a converged final state, from which
a time-evolved signal is constructed and Fourier transformed.
Figure~\ref{figS6}(a)-(c) show the two-dimensional Fourier-amplitude
maps (color represents $\log_{10}|\text{amplitude}|$) for the initial
states $\ket{00000001}$, $\ket{00000010}$, and $\ket{00000101}$, respectively,
plotted against frequency $\omega/J$ (horizontal axis) and normalized
evolution time $J\tau$ (vertical axis). Distinct bright stripes appear at
frequencies corresponding to the four lowest eigenenergy levels $E_0$, $E_1$, $E_2$, and $E_3$.

These maps also help distinguish true neighboring eigenstates from
mere peak broadening. For example, the bright region near $E_0$ shows
adjacent pixels with correlated amplitude variations across $\tau$,
characteristic of a single broadened peak. In contrast, the bright
regions for $E_2$ and $E_3$ exhibit uncorrelated amplitude fluctuations
as $\tau$ changes, confirming that they arise from two distinct, closely
spaced eigenstates rather than from the broadening of a single peak.
Thus, a rough $\tau$ scan already reveals the approximate energy levels
and their ordering.

The choice of $\tau$ for high-precision fitting is guided by the amplitude
dependence on $\tau$ (see Fig. 3(h) in the main text). The amplitude
oscillates rapidly with $\tau$; selecting a $\tau$ near a local maximum
provides a high signal-to-noise ratio for the subsequent multi-frequency
fitting of that level. Sharp drops in amplitude correspond to $\tau$ values
for which the optimization converges to other eigenstates (sometimes at
much higher energies) or yields a poorly converged final state.
In practice, for the initial state $\ket{00000001}$, the $\tau$ used in the
main text ($J\tau = 13.2$) is not at the absolute maximum for $E_0$ alone,
but at a value that yields simultaneously large amplitudes for both $E_0$ and $E_2$,
enabling the simultaneous extraction of these two levels as shown in
Fig. 4(e) and 4(f) of the main text. While a single ($\ket{\psi_0},\tau)$
configuration could in principle be used to extract all four lowest levels,
we employed three different initial states with $\tau$ values selected near
the respective amplitude maxima for $E_0\&E_2$, $E_1$, and $E_3$ to
demonstrate the flexibility and robustness of the protocol.

Supplementary Figure~\ref{figS6}(d)-(f) present the corresponding two-dimensional
amplitude maps obtained from simulation for the 8-qubit linear-chain Hamiltonian
and circuit. Although the amplitude peaks are somewhat sparser---especially
in panel (f), indicating a higher probability for the optimization to
converge to higher-energy eigenstates---clear maxima are still observable,
and the selected $\tau$ values yield final states of comparable fidelity.
This confirms that the simplified linear connectivity does not degrade
the performance of the variational optimization, and the $\tau$-selection
procedure remains effective across different hardware graphs.

\begin{figure}[htbp]
    \centering
    \includegraphics[width=0.95\linewidth]{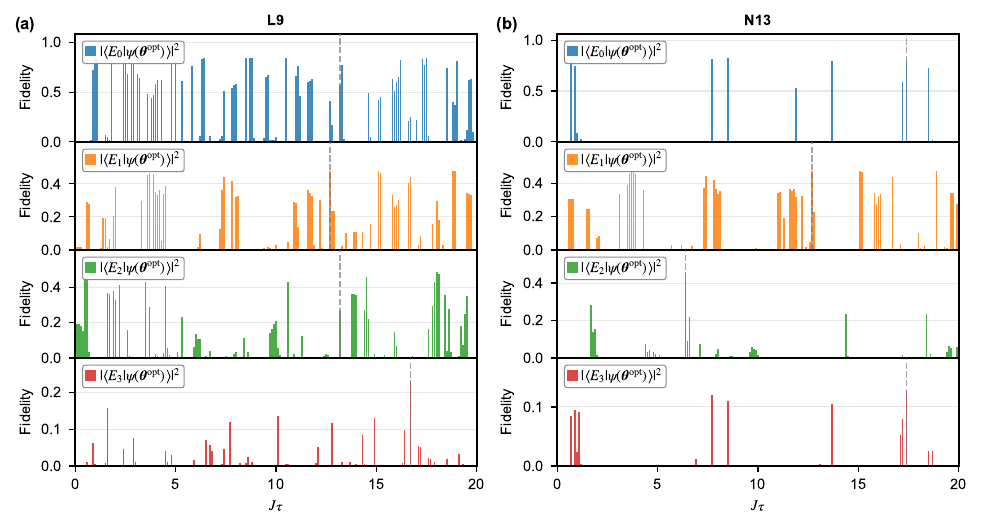}
    \caption{State-composition analysis of optimized final states for
    varying evolution times $\tau$. (a) Results for the 8-qubit cube-connected
    L9 processor. (b) Corresponding results for the 8-qubit linear chain on
    the 13-qubit processor. In each panel, the four sub-panels (top to bottom)
    show the squared overlap (fidelity) of the optimized final state with
    the four lowest eigenstates $E_0$--$E_3$ as a function of the normalized
    evolution time $J\tau$. A high overlap indicates a larger proportion of
    the corresponding eigenstate in the final prepared state. The variation of
    these overlaps with $J\tau$ aligns well with the $\tau$-dependent Fourier
    amplitude shown in Fig.~\ref{figS6}, confirming the consistency between the
    time-domain spectral analysis and the direct eigenstate-overlap
    measurement.}
    \label{figS7}
\end{figure}

Supplementary Figure~\ref{figS7} provides a direct, state-overlap verification
of the eigenstate composition of the optimized final states shown in Fig.~\ref{figS6}.
For each initial state and each normalized evolution time $J\tau$, we compute
the squared overlap (fidelity) of the optimized final state with the four
lowest eigenstates $E_0$--$E_3$. The resulting fidelity maps (Fig.~\ref{figS7}(a)
for the L9 cube and Fig.~\ref{figS7}(b) for the N13 linear chain)
show how the proportion of each eigenstate in the final state varies with $J\tau$.
High overlap in a given sub-panel indicates that the final state is rich in that
particular eigenstate. The oscillation patterns of these overlaps with $J\tau$ match
closely the $\tau$-dependent Fourier amplitudes displayed in Fig.~\ref{figS6},
confirming that the time-domain spectral analysis reliably reflects the actual
eigenstate composition of the prepared states. This consistency validates the
use of Fourier amplitude as a proxy for selecting $\tau$ values that maximize
the signal of a target eigenlevel for subsequent precision fitting.

\makeatletter
\let\addcontentsline\@gobblethree
\makeatother

\end{document}